\documentclass[aps,prl,twocolumn,groupedaddress,showpacs]{revtex4}
\usepackage{epsfig,graphics}
\usepackage{graphicx}

\begin{document}

\title{Estimation of Purcell factor from mode-splitting spectra in an optical microcavity}

\author{\c{S}ahin Kaya \"{O}zdemir}
\email{ozdemir@ese.wustl.edu}
\affiliation{Department of Electrical and Systems Engineering, Washington University, St. Louis, Missouri 63130, USA}
\author{Jiangang Zhu}
\email{jzhu@seas.wustl.edu}
\affiliation{Department of Electrical and Systems Engineering, Washington University, St. Louis, Missouri 63130, USA}
\author{Lina He}
\affiliation{Department of Electrical and Systems Engineering, Washington University, St. Louis, Missouri 63130, USA}
\author{Lan Yang}
\email{yang@seas.wustl.edu}
\affiliation{Department of Electrical and Systems Engineering, Washington University, St. Louis, Missouri 63130, USA}

\begin{abstract}
We investigate scattering process in an ultra-high-$\emph{Q}$ optical microcavity coupled to subwavelength scatterers by introducing {\it splitting quality} $\emph{Q}_{\rm sp}$, a dimensionless parameter defined as the ratio of the scatterer-induced mode splitting to the total loss of the coupled system. A simple relation is introduced to directly estimate the Purcell factor from single-shot measurement of transmission spectrum of scatterer-coupled cavity. Experiments with polystyrene (PS) and gold (Au) nanoparticles, Erbium ions and Influenza A virions show that Purcell-factor-enhanced preferential funneling of scattering into the cavity mode takes place regardless of the scatterer type. Experimentally determined highest $\emph{Q}_{\rm sp}$ for single PS and Au nanoparticles are 9.4 and 16.19 corresponding to Purcell factors with lower bounds of 353 and 1049, respectively. The highest observed $\emph{Q}_{\rm sp}$ was 31.2 for an ensemble of Au particles. These values are the highest $\emph{Q}_{\rm sp}$ and Purcell factors reported up to date.
\end{abstract}


\maketitle

There has been a growing interest in ultra-high-$\emph{Q}$ optical whispering gallery mode (WGM) microcavities as they provide a suitable platform to study both fundamental physical phenomena (e.g., opto-mechanics, light-matter interactions) and to fabricate practical devices (e.g., single molecule and nanoparticle detection, optical sensors, etc.) \cite{Vahala}. Observed, for the first time, by Il'chenko {\it et al.} \cite{Gorodetsky} and later addressed comprehensively by Weiss {\it et al.} \cite{Weiss}, mode-splitting in WGM microcavities have developed into a useful tool to investigate light-matter interaction \cite{Mazzei,Kippenberg1,Jiangang1,Jiangang2,Kippenberg2,Borselli,Grudinin}.

Mode splitting occurs due to the lifting of the degeneracy of the clockwise (CW) and the counter-clockwise (CCW) propagating 
WGMs as a result of their coupling via scattering centers intrinsically present (i.e., due to material inhomogeneity, structural defects, dust contaminations) or intentionally introduced (e.g., particle depositions, fiber tips, etc.) into the microcavity mode volume. Superposition of CW and CCW modes form two orthogonal standing wave modes (SWMs), one of which experiences a frequency shift and extra linewidth broadening with respect to the other. This leads to a doublet in the transmission spectrum.

Previously, we used mode-splitting in passive and active microcavities to measure individual nanoparticles \cite{Jiangang1,Lina}. It was also used to demonstrate Purcell-factor-enhanced scattering by considering a collective scattering rate of an ensemble of silicon nanocrystals embedded randomly into a microcavity \cite{Kippenberg2}. These results were interesting; however they do not provide insight into the quantification of mode splitting and the Purcell-factor-enhanced scattering at the level of a single scatterer.

Purcell effect takes place due to light scattering regardless of whether the scatterer is quantum (modeled as discrete energy levels) or classical (modeled as a dipole). It is an important figure of merit for quantifying the ability of a cavity to couple to an emitter or a scatterer. Thus, it is important to develop reliable and easily accessible methods to measure Purcell factor accurately. In this Letter, we report the direct estimation of Purcell factor from mode splitting in the transmission spectrum with a single-shot measurement, and clarify the role of individual scatterer on the quality of mode-splitting and hence on Purcell-enhanced Rayleigh scattering.

The proposed method for direct estimation of Purcell factor and demonstrated high Purcell-enhanced scattering efficiencies at single scatterer resolution, using PS and Au nanoparticles, and Influenza A (InfA) virions, will play an important role in detecting and studying classical and quantum scatterers independent of their type and internal structures. For example, quantum scatterers such as atoms and molecules can be detected by mode splitting spectra and Purcell-enhanced scattering in a microcavity with a resonance largely detuned from their optical transitions such that they are not optically excited. This will provide an alternative technique to fluorescence, absorption and ionization based atom detection and will eliminate the need for near-resonant lasers.

The amount of mode-splitting and the difference in the linewidths of the individual resonances in the doublet are given as $2g=-\alpha f^2 (r)\omega/V$ and $2\Gamma=-2g\alpha\omega^3/3\pi c^3$ where the polarizability $\alpha$ is defined as $\alpha=4\pi R^3 (n^2-1)/(n^2+2)$ for a single particle of radius $R$ and refractive index $n$, $V$ denotes the mode volume, $f(r)$ the normalized mode distribution, $c$ the speed of light, and $\omega$ the angular frequency of the initial resonant WGM \cite{Mazzei,Jiangang1}. The coupling strength of the initially degenerate WGMs can be described with the {\it mode-splitting quality} $\emph{Q}_{\rm sp}$ as
\begin{equation}\label{N01}
\emph{Q}_{\rm sp}=\frac{2|g|}{\Gamma+(\omega/\emph{Q})}
\end{equation} where $\omega/\emph{Q}$ is the initial resonance linewidth. To resolve the mode-splitting in the transmission spectra, $\emph{Q}_{sp}>1$ should be satisfied, i.e., $2|g|$ should exceed the mean of intrinsic and scatterer induced decay rates of the system:
\begin{equation}\label{N02}
2|g|>\Gamma+(\omega/\emph{Q}).
\end{equation}
If the scattering losses dominate the cavity losses, i.e., $\Gamma\gg(\omega/\emph{Q})$, Eq. (\ref{N01}) simplifies to $\emph{Q}_{\rm sp}^{\rm I}=2|g|/\Gamma$. Then quality of mode-splitting can be related to the nanoparticle through particle polarizability as
\begin{equation}\label{N03}
\emph{Q}_{\rm sp}^{\rm I}=\left|\frac{2g}{\Gamma}\right|=\frac{3\lambda^3}{4\alpha\pi^2}.
\end{equation}
Thus, in this regime mode-splitting quality is affected neither by the $\emph{Q}$ of the cavity modes nor by $f(r)$. It is determined by the resonance wavelength of the initial WGM and the particle polarizability.

On the other hand, if the cavity losses dominates, i.e., $\Gamma\ll(\omega/\emph{Q})$, Eq. (\ref{N01}) simplifies to
\begin{equation}\label{N04}
\emph{Q}_{\rm sp}^{\rm II}=\frac{2|g|\emph{Q}}{\omega}=\frac{\alpha f^2(r)\emph{Q}}{V}\leq \frac{\alpha\emph{Q}}{V}
\end{equation} where the upper bound is obtained by setting $f^2(r)=1$. Thus, in this regime $\emph{Q}_{\rm sp}$ increases linearly with particle polarizability and the quality factor of the WGM. Moreover, the mode field distribution and volume strongly affect the splitting-quality in this regime. Since one does not have precise control on the location of the nanoparticle within the microcavity mode volume, $f(r)$ is unknown and it plays a limiting role for $\emph{Q}_{\rm sp}$ in this regime.
\begin{figure}\epsfxsize=5.5cm \epsfbox{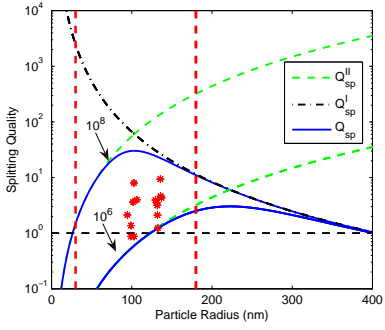} \caption{Splitting quality versus particle size ($R$) for PS (refractive index $n=1.59$) particles for initial cavity quality factor $\emph{Q}$ equals to $10^6$ and $10^8$. Vertical red lines denotes the size range of particles used in the experiments. Mode splitting is resolved if $\emph{Q}_{\rm sp}$ lies above the horizontal dashed line ($\emph{Q}_{\rm sp}>1$). The points labeled with $\star$ denote the $\emph{Q}_{\rm sp}$ values obtained in single-particle experiments with particles of mean radii $R=100 \rm{nm}$ and $R=135\rm{nm}$.}\label{fig1}\end{figure}
The expressions obtained for splitting quality for {\it scattering losses dominant} and for {\it cavity losses dominant} regimes in Eqs. (\ref{N03}) and (\ref{N04}), respectively, also establish the relationship for the lower bound of Purcell factor ${\mathcal F}$:
\begin{equation}\label{N05}
{\mathcal F}\geq \emph{Q}_{\rm sp}^{\rm I}\emph{Q}_{\rm sp}^{\rm II}\geq 4\emph{Q}_{\rm sp}^2
\end{equation} where ${\mathcal F}=(3/4\pi^2)\lambda^3(\emph{Q}/V)$. The second part of the inequality is obtained using $1/\emph{Q}_{\rm sp}=1/\emph{Q}_{\rm sp}^{\rm I}+1/\emph{Q}_{\rm sp}^{\rm II}$ and $4\emph{Q}_{\rm sp}^2\leq\emph{Q}_{\rm sp}^{\rm I}\emph{Q}_{\rm sp}^{\rm II}$. The relation in Eq. (\ref{N05}) can be used to estimate the lower bound of Purcell factor in an experiment by using the $\emph{Q}_{\rm sp}$ obtained from transmission spectrum of single-particle coupled microcavity.

In Fig. \ref{fig1}, we depict the dependence of $\emph{Q}_{\rm sp}$ on the particle size and the quality-factor of the microcavity. For particles larger than a critical size, we have $\emph{Q}_{\rm sp}\sim\emph{Q}_{\rm sp}^{\rm I}$ suggesting that the additional decay channel formed by the particle dominates the cavity losses. On the other hand, for particles smaller than a critical size, we see $\emph{Q}_{\rm sp}\sim\emph{Q}_{\rm sp}^{\rm II}$ which suggests that small particles do not induce significant loss when compared to the cavity losses. In between these two critical particle sizes, there is a region where the splitting-quality can be explained neither by $\emph{Q}_{\rm sp}^{\rm I}$ nor by $\emph{Q}_{\rm sp}^{\rm II}$; hence the exact expression of $\emph{Q}_{\rm sp}$ given in Eq.(\ref{N01}) should be used. Furthermore, we see that for a given $\emph{Q}$, there is a critical lower and critical upper bound for the particle size beyond which $\emph{Q}_{\rm sp}$ becomes less than one. While the upper bound is determined by particle size, the lower bound is strongly affected by $\emph{Q}$. Therefore, by choosing a mode with higher $\emph{Q}$, mode-splitting induced by particles with radii of a few nanometers can be resolved, subsequently such particles can be detected and measured. It is worth noting that $\emph{Q}_{\rm sp}$ for a fixed cavity $\emph{Q}$-factor is upper bounded by $\emph{Q}_{\rm sp}^{\rm I}=\emph{Q}_{\rm sp}^{\rm II}$, i.e., $\omega/\emph{Q}=\Gamma$ implying $\emph{Q}_{\rm sp}=\emph{Q}_{\rm sp}^{\rm I}/2=\emph{Q}_{\rm sp}^{\rm II}/2$, and the particle size which pushes $\emph{Q}_{\rm sp}$ to its maximum value becomes smaller with increasing cavity $\emph{Q}$-factor. One can easily show that the particle maximizing $\emph{Q}_{\rm sp}$ should have polarizability satisfying $\alpha=3\pi^{-2}f(r)^{-1}\lambda^3\sqrt{{\mathcal F}}/4$.

\begin{figure}\epsfxsize=5cm \epsfbox{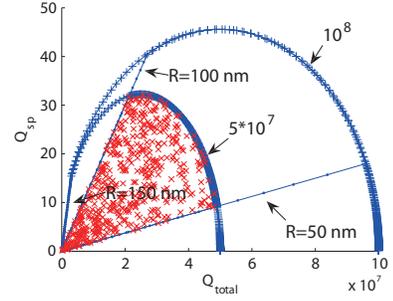} \caption{Total quality factor $\emph{Q}_{\rm T}$ versus mode splitting quality $\emph{Q}_{\rm sp}$ for various values of PS particle size $R$. Red $\star$ points correspond to randomly chosen $(\emph{Q}_{\rm T},R)$ pairs in the region $10^6\leq\emph{Q}\leq 5\times10^7$ and $50~{\rm nm}\leq R\leq 100~{\rm nm}$. Linear boundaries are obtained by varying $\emph{Q}$ in $10^6\leq\emph{Q}_{\rm T}\leq 10^8$ for fixed $R$, and inverted parabola-like boundaries are obtained by decreasing $R$ down to $10~{\rm nm}$ for fixed $\emph{Q}$.}\label{fig2}\end{figure}
Here we discuss how the mode-splitting quality $\emph{Q}_{\rm sp}$ depends on the total quality factor $\emph{Q}_{\rm T}$ of a scatterer-coupled microcavity (Fig.\ref{fig2}). $\emph{Q}_{\rm T}$ takes into account both the cavity losses and particle related losses, i.e.,$\emph{Q}_{\rm T}=\omega[\Gamma+(\omega/Q)]^{-1}$. Then $\emph{Q}_{\rm sp}=\emph{Q}_{\rm T}(2g/\omega)$ implying that $\emph{Q}_{\rm sp}$ increases linearly with $\emph{Q}_{\rm T}$ with a slope of $2g/\omega$ if $R$ (or $\alpha$) is kept constant, i.e., $g$ and $\Gamma$ are fixed too. When $\emph{Q}$ is very small, process is dominated by cavity losses. Since scatterer-related-losses are constant (i.e., $R$ is fixed), increasing $\emph{Q}$ decreases cavity losses. When cavity and scatterer losses are equal ($\Gamma=\omega/\emph{Q}$), we have $\emph{Q}_{\rm T}=\omega/2\Gamma$ and $\emph{Q}_{\rm sp}=\emph{Q}_{\rm sp}^{\rm I}/2=g/\Gamma$. If $\emph{Q}$ is increased beyond this value, cavity losses becomes smaller than the scatterer losses. In the limit of large $\emph{Q}$, we have $\emph{Q}_{\rm T}=\omega/\Gamma$ leading to $\emph{Q}_{\rm sp}=\emph{Q}_{\rm sp}^{\rm I}=2g/\Gamma$. Thus 
two sections can be defined: (L1) Cavity losses dominant section ($\omega/\emph{Q}>\Gamma$) which is the part of the linear curve within $0\leq\emph{Q}_{\rm T}<\omega/2\Gamma$ and $0\leq\emph{Q}_{\rm sp}<\emph{Q}_{\rm sp}^{\rm I}/2$. (L2) Scattering dominant section ($\omega/\emph{Q}<\Gamma$) is the part bounded with $\emph{Q}_{\rm T}>\omega/2\Gamma$ and $\emph{Q}_{\rm sp}^{\rm I}/2<\emph{Q}_{\rm sp}\leq\emph{Q}_{\rm sp}^{\rm I}$.
\begin{figure}\epsfxsize=6cm \epsfbox{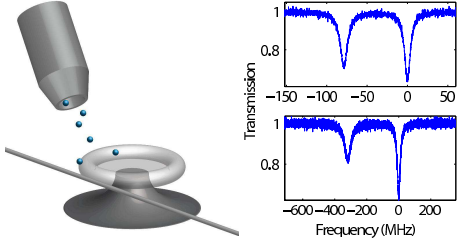} \caption{Schematics of a tapered fibre interfaced to a silica microtoroid with deposited nanoparticles, and typical transmission spectra captured with a photo detector for PS particles of radii $R=100 {\rm nm}$ (top) and $R=135 {\rm nm}$ (bottom).}\label{fig3}\end{figure}

Next we investigate what happens if $\emph{Q}$ is kept constant at $\emph{Q}_0$ and the particle size $R$ is decreased from its initial value of $R_0$ (Fig.\ref{fig2}). If the process is initially in the region $\omega/\emph{Q}_0>\Gamma_{R_0}$, decreasing particle size will significantly decrease $\Gamma$ due to its $\Gamma\propto R^6$ dependence which makes it much lower than $\omega/\emph{Q}_0$. Due to $g\propto R^3$ dependence, $g$ also decreases but at a slower pace. Then, in this case the process continues to be dominated by cavity losses, and $\emph{Q}_{\rm sp}$ decreases with decreasing $R$. If the process is initially in the region $\omega/\emph{Q}_0<\Gamma_{R_0}$, the evolution is strongly affected by $g$ and $\Gamma$ rather than $\omega/\emph{Q}$. Thus, the slower pace of decrease in $g$ compared to that in $\Gamma$ allows an increase in $\emph{Q}_{\rm sp}$. This continues until the point of $\omega/\emph{Q}=\Gamma_{R}$ and $\emph{Q}_{\rm sp}=g/\Gamma$. Further decrease of $R$ shifts the process from scatterer-dominated regime to cavity-loss-dominated regime in which the effect of $\omega/\emph{Q}$ is much stronger. Thus, after stepping into this regime, the process will stay in this regime with decreasing $R$ until $\emph{Q}_{\rm sp}$ decreases from its peak value to zero.

Figure \ref{fig3} shows transmission spectra obtained for single PS particles of $R=100 {\rm nm}$ and $135 {\rm nm}$ coupled to a microtoroid \cite{Jiangang1}. For these spectra, we calculate $\emph{Q}_{\rm sp}=8$ and $\emph{Q}_{\rm sp}=9.4$ from Eq. (\ref{N01}). Then Eq. (\ref{N05}) yields the corresponding Purcell factors as ${\mathcal F}\geq 256$ and ${\mathcal F}\geq 353$. This is the best Purcell factor obtained for a single PS nanoparticle. The $\star$ points in Fig.\ref{fig1} denote experimentally obtained $\emph{Q}_{\rm sp}$ for different depositions of nanoparticles whose sizes were estimated using $\alpha=(3\lambda^3/8\pi^2)(1/\emph{Q}_{\rm sp}^{\rm I})$ \cite{Jiangang1}. Different $\emph{Q}_{\rm sp}$ values for the same estimated size are due to different $f(r)$ at each deposition.

For multi-scatterer case, SWMs are affected by all particles in varying degrees and they are spectrally-shifted from the original WGM resonance. We denote $(\omega_j,\gamma_j)$ with $j=0,1,2$ as the angular resonance frequencies and linewidths of the initial (pre-scatterer) WGM ($j=0$), and low ($j=1$) and high ($j=2$) frequency SWMs created after the deposition of $N$ scatterers. Then, we have
\begin{eqnarray}
\omega_j-\omega_0&=&\sum_{i=1}^{N}2g_i\sin^2(\phi_i+j\frac{\pi}{2})\label{N06}\\
\gamma_j-\gamma_0&=&\sum_{i=1}^{N}2\Gamma_i\sin^2(\phi_i+j\frac{\pi}{2})\label{N07}
\end{eqnarray}
where $\phi_i$ denotes the spatial distance of $i$-th particle from the anti-node of a SWM. From Eq. (\ref{N06}), we find the amount of mode splitting as
\begin{equation}\label{N08}
\chi=\omega_2-\omega_1=\sum_{i=1}^{N}2g_i\cos(2\phi_i).
\end{equation} Moreover, Eq. (\ref{N07}) yields the linewidth relation as
\begin{equation}\label{N09}
\gamma_1+\gamma_2=2\gamma_0+\sum_{i=1}^{N}2\Gamma_i.
\end{equation} To resolve the split modes after the deposition of $N$ scatterers, $|\chi|>(\gamma_1+\gamma_2)/2$ should be satisfied. Thus, mode splitting quality for $N$-scatterer then becomes
\begin{equation}\label{N10}
\emph{Q}_{\rm sp}^{(N)}=\frac{2|\chi|}{\gamma_1+\gamma_2}=\frac{2\left|\sum_{i=1}^{N}g_i\cos(2\phi_i)\right|}{(\omega_0/Q)+\sum_{i=1}^{N}\Gamma_i}
\end{equation}  which has a form similar to Eq. (\ref{N01}). Assuming that the particles have the same polarizability and using triangle inequality and $0\leq|\cos(2\phi_i)|\leq 1$, we find that splitting quality in the limit of scatterer dominant regime becomes $\emph{Q}_{\rm sp}^{(N)I}\leq\emph{Q}_{\rm sp}^{I}$ where $\emph{Q}_{\rm sp}^{I}$ is given in Eq. (\ref{N03}). If we further set $f^2(r_i)=f^2(r)$ and consider the cavity losses dominant regime, mode splitting quality becomes $\emph{Q}_{\rm sp}^{(N)II}=\alpha f^2(r)\emph{Q}N/V\leq\alpha \emph{Q}N/V$ where in the last part we assumed maximum overlap between the WGM and each of the scatterers, i.e.,$f^2(r)=1$. Assuming that all particles have the same $\alpha$, and defining $g=g'f^2(r_i)$ and $\Gamma=\Gamma'f^2(r_i)$ with $g'=-\alpha\omega/2V$ and $\Gamma'=\alpha^2\omega^4/6\pi Vc^3$, it is easy to see that
\begin{eqnarray}\label{N11}
\emph{Q}_{\rm sp}^{(N)}\leq \frac{2|g'|\sum_{i=1}^{N}f^2(r_i)}{(\omega_0/Q)+\Gamma'\sum_{i=1}^{N}f^2(r_i)}
\leq\frac{2N|g'|}{(\omega_0/Q)+N\Gamma'}
\end{eqnarray} where we used $f_i^2(r)=f^2(r)=1$. For $N=1$, Eq. (\ref{N11}) reproduces the results for the single scatterer case. Following the procedure carried out for a single scatterer, we find the Purcell factor for multi-scatterer case as
\begin{equation}\label{N12}
{\mathcal F}^{(N)}\geq 4N^{-1}\emph{Q}_{\rm sp}^{(N)2}
\end{equation}
from which one can deduce that the minimum attainable Purcell factor decreases with increasing number of scatterers due to the increasingly dominating scatterer induced dissipation. Thus lower bound of the Purcell factor can be estimated from the experimentally obtained transmission spectra and using the calculated $\emph{Q}_{\rm sp}^{(N)}$ in Eq. (\ref{N12}) provided that the number of particles $N$ is known.

\begin{figure}\epsfxsize=6cm \epsfbox{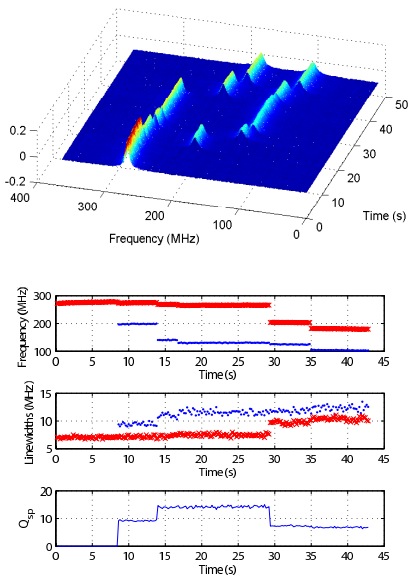} \caption{A series of transmission spectra obtained for consecutive depositions of PS nanoparticles ($R=135 {\rm nm}$) into the mode volume of the resonator and the corresponding amount of mode splitting, additional linewidth broadening and the mode splitting quality. A total of five nanoparticle deposition is depicted. Each frequency jump in the spectra corresponds to a single nanoparticle deposition event.}\label{fig4}\end{figure}

Figure \ref{fig4} shows the evolution of transmission spectra and calculated mode splitting, additional linewidth broadening and mode splitting quality as nanoparticles are deposited into the resonator mode volume. Each discrete jump corresponds to deposition of a single nanoparticle. With each particle deposition, mode splitting is either enhanced or reduced depending on the location of the particle on the mode volume. Measured values of $\emph{Q}_{\rm sp}^{(1)}=9.2$, $\emph{Q}_{\rm sp}^{(2)}=14.1$, $\emph{Q}_{\rm sp}^{(3)}=14.0$, $\emph{Q}_{\rm sp}^{(4)}=7.3$ and $\emph{Q}_{\rm sp}^{(5)}=6.8$, respectively, corresponds to ${\mathcal F}^{(1)}\geq 338$, ${\mathcal F}^{(2)}\geq 397$, ${\mathcal F}^{(3)}\geq 261$, ${\mathcal F}^{(4)}\geq 53$, and ${\mathcal F}^{(5)}\geq 36$. Estimated Purcell factor for two-particle event, which is also the highest Purcell factor ever reported \cite{Kippenberg2, Pitanti}, implies that at least $99.75\%$ of the scattered light is captured by the doubly degenerate cavity modes \cite{Kippenberg2}.

We investigated the effect of scatterer properties with experiments using Au nanoparticles, Erbium (${\rm Er}^{+3}$) ions and InfA virions. For a single Au nanoparticle ($R=100 {\rm nm}$), we measured $\emph{Q}_{\rm sp}=16.2$ corresponding to ${\mathcal F}^{(1)}\geq 1049$. With subsequent depositions, $\emph{Q}_{\rm sp}$ changed as $\emph{Q}_{\rm sp}^{(2)}=20.16$, $\emph{Q}_{\rm sp}^{(3)}=24.64$ and $\emph{Q}_{\rm sp}^{(4)}=20.88$ with the estimated ${\mathcal F}^{(2)}\geq 812$, ${\mathcal F}^{(3)}\geq 809$, and ${\mathcal F}^{(4)}\geq 435$. For an ensemble of $N\leq 20$, the highest measured $\emph{Q}_{\rm sp}$ was $31.2$ with the resonance modes shifted from each other by $2|g|=364.9 {\rm MHz}$ with a mean linewidth of $11.7 {\rm MHz}$. This corresponds to ${\mathcal F}\geq 194.7$ implying a capture efficiency higher than $99.5\%$. The splitting quality of $\emph{Q}_{\rm sp}=31.2$ is the highest reported up to date \cite{Kippenberg2}. Next, we performed experiments using Er-doped silica microtoroids \cite{Lina}. Gain provided by optically pumping ${\rm Er}^{+3}$ ions enables detection of small intrinsic mode splitting, which could not be observed before optical pumping. Such small mode splittings yielded $\emph{Q}_{\rm sp}\sim3.5$. Finally, we deposited InfA virions one-by-one on a microtoroid. Estimated mode splitting qualities are $\emph{Q}_{\rm sp}^{(1)}=2.5$, $\emph{Q}_{\rm sp}^{(2)}=5.9$, $\emph{Q}_{\rm sp}^{(3)}=9.8$, $\emph{Q}_{\rm sp}^{(4)}=9.2$ and $\emph{Q}_{\rm sp}^{(5)}=12.6$, respectively corresponding to ${\mathcal F}^{(1)}\geq 25$, ${\mathcal F}^{(2)}\geq 69$, ${\mathcal F}^{(3)}\geq 128$, ${\mathcal F}^{(4)}\geq 84$, and ${\mathcal F}^{(5)}\geq 127$. These results demonstrate that preferential backscattering into the resonator takes place regardless of the scatterer type, and that it can be quantified from the mode splitting spectrum with a single-shot measurement.

In conclusion, we have investigated scattering induced mode splitting in ultra-high-$\emph{Q}$ microcavities, and introduced a formalism to quantify the quality of mode splitting for single and multiple-scatterers as the particles enter the mode volume of the resonator. This formalism enables estimating the lower bound for Purcell factor and Rayleigh scattering capture efficiency from the transmission spectra. This will help understanding the Purcell-factor-enhanced scattering process at the single particle level, and boosting the sensitivity of detecting individual nanoparticles entering to or embedded into the mode volume of a resonator. Finally, this method can be further extended for a wide class of microcavities with embedded, adsorbed or deposited emitters and scatterers (e.g., nanocrystals, plasmonic particles, atoms, molecular gases) independent of their internal structures using collective or individual enhancement effects, thus providing a non-invasive detection scheme.

The authors gratefully acknowledge the support from NSF (Grant No. 0954941). This work was performed in part at the NRF/NNIN (NSF award No. ECS-0335765) of Washington University in St. Louis.

\end{document}